\documentclass[twocolumn,showpacs,preprintnumbers,amsmath,amssymb]{revtex4}

\usepackage{graphicx}
\usepackage{dcolumn}
\usepackage{bm}

\newcommand{\bea}{\begin{eqnarray}} 
\newcommand{\beq}{\begin{equation}} 
\newcommand{\ear}{\end{array}} 
\newcommand{\eea}{\end{eqnarray}} 
\newcommand{\eeq}{\end{equation}}

\begin{document}

\title{\Large Superdeformation and hyperdeformation in the $^{108}$Cd 
nucleus.}

\author{A.\ V.\ Afanasjev$^{(1,2)}$, S.\ Frauendorf$^{(1,3)}$} 

\address{$^{1}$Department of Physics, University of Notre Dame, Notre Dame, 
Indiana 46556, USA}

\address{$^{2}$ Department of Physics and Astronomy, Mississippi State
University, MS 39762, USA} 

\address{$^{3}$IKH, Research Center Rossendorf, Dresden, Germany}

\date{\today}

\begin{abstract}
The superdeformation and hyperdeformation in $^{108}$Cd have been 
studied for the first time within the framework of the fully 
self-consistent cranked mean field theory, namely, cranked relativistic 
mean field theory. The structure of observed superdeformed bands
1 and 2 have been analyzed in detail. The bumps seen in their dynamic 
moments of inertia are explained as arising from unpaired band
crossings. This is contrary to an explanation given earlier within
the framework of projected shell model. It was also concluded that 
this nucleus is not doubly magic SD nucleus.
\end{abstract}
 

\maketitle

  Since the discovery of superdeformation in $^{152}$Dy two decades
ago \cite{Dy152}, nuclear super- and hyperdeformation have been in 
the focus of attention of nuclear structure community. At present, 
superdeformation (SD) has been discovered in different mass regions 
and extensively studied experimentally \cite{SD-sys} and theoretically 
(see, for example, Ref.\ \cite{A150} and references therein). On the 
other side, nothing is known experimentally about hyperdeformation 
(HD) apart from some indications of this phenomenon at low spin in 
the uranium nuclei \cite{K.98}.

   Recent observation of the very extended shapes in $^{108}$Cd 
\cite{Cd108-1,Cd108-2} has opened new region of superdeformation 
and renewed interest to the study of hyperdeformation \cite{HD}. 
Two SD 
bands have been found in this nucleus. In the present manuscript, 
they are labeled according to Ref.\ \cite{Cd108-2}, namely, the 
lowest SD band as band 1 and the excited SD band as band 2.
These experiments were strongly motivated by earlier calculations 
of Ref.\ \cite{WD.95} and more recent studies of Ref.\ \cite{C.01}.

So far theoretical interpretation has been performed only for band 
1 within the framework of projected shell model (PSM) (Ref.\ \cite{PSM}). 
Clear deficiency of this description is the fact that the equilibrium 
deformation is not defined within some ``variational'' 
procedure, but is adjusted for a better description of moments of 
inertia. It was concluded in Ref.\ \cite{PSM} that the low-$K$ 
$i_{13/2}$ proton orbitals are responsible for the observed bump 
in the dynamic moment of inertia $J^{(2)}$ at low rotational frequencies 
and that the 2-quasiparticle configurations from these orbitals 
dominate the structure of the observed states. However, it is 
well known that the pairing is considerably quenched at high 
rotational frequencies and superdeformation \cite{SGBGV.89,CNS}. 
For example, no paired band crossing is observed above rotational 
frequencies $\omega \sim 0.5$ MeV in the $A\sim 150$ \cite{A150,A150-sp,VRAL} 
and $\omega \sim 1.0$ MeV in the $A\sim 60$ \cite{A60,VRAL} mass 
regions of superdeformation and the experimental data above these 
frequencies are well (within $\approx 5\%$ accuracy for the 
moments of inertia) described by the cranked relativistic mean
field (CRMF) theory with no pairing. Considering that the $A\sim 110$ 
mass region of superdeformation is located between these two regions, 
it is reasonable to expect that the influence of pairing will be 
negligible above $\omega \geq 0.8-0.9$ MeV. Thus, rephrasing the 
results of the PSM studies in Ref.\ \cite{PSM} to the language of 
the calculations without pairing, one can conclude that the 
$\pi 6^2$ configuration has to be assigned to the band 1 
in $^{108}$Cd above the band crossing. 
The $\pi 6^2$ SD configurations are active in the medium-mass part of 
the rare-earth
region of superdeformation [Gd ($Z=64$) nuclei] \cite{A150}. 
However, the Fermi level is located at much lower $Z=48$ in $^{108}$Cd 
as compared with the $A\sim 150$ mass region of superdeformation, and, 
thus, considerably larger deformation and higher rotational frequencies 
than in the rare-earth region of SD will be required to have the two 
lowest $N=6$ proton orbitals occupied.

   In the present manuscript, the CRMF theory \cite{KR.89,A150} is used 
for detailed investigation 
of the structure of observed SD bands in $^{108}$Cd and the nature 
of hyperdeformation in the $A\sim 110$ mass region. Additional insight 
has been provided by the cranked Nilsson-Strutinsky (CNS \cite{CNS}) 
calculations performed with the standard Nilsson parameters \cite{BR.86}. 
In both calculations, the pairing is neglected. The CRMF equations are 
solved in the basis of an anisotropic three-dimensional harmonic 
oscillator in Cartesian coordinates with the deformation parameters 
$\beta_0=0.65$, $\gamma=0^{\circ}$ and oscillator frequency
$\hbar \omega_0=41 A^{-1/3}$ MeV. All fermionic and bosonic states 
belonging to the shells up to $N_F=14$ and $N_B=16$ are taken into 
account in the diagonalization of the Dirac equation and the matrix 
inversion of the Klein-Gordon equations, respectively. The detailed 
investigation indicates that this truncation scheme provides good 
numerical accuracy. The NL1 set \cite{NL1} is used for the RMF 
Lagrangian. As follows from our experience \cite{A150-sp,A250}, this 
set provides reasonable description of the single-particle energies.

\begin{figure}[ht]
\includegraphics[angle=0,width=8cm]{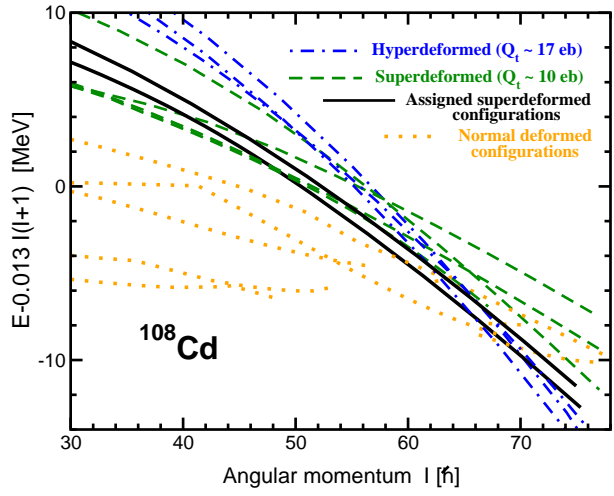}
\caption{\label{eld}  
Energies of the configurations calculated in the 
CRMF theory relative to a smooth liquid drop reference $A I(I+1)$, with 
the inertia parameter $A=0.013$. 
}
\end{figure}

  The results of the CRMF calculations for the configurations forming 
the yrast line or located close to it in energy are shown in Fig.\ 
\ref{eld}. According to these calculations, normal-deformed bands, 
many of which show high triaxiality indicative of approaching band 
termination, dominate the yrast line up to $I \approx 68\hbar$. At 
higher spin hyperdeformed bands become yrast. The SD bands 
are never yrast, but come close to the yrast line at 
$I\approx 66\hbar$.

\begin{figure}[ht]
\includegraphics[angle=0,width=6cm]{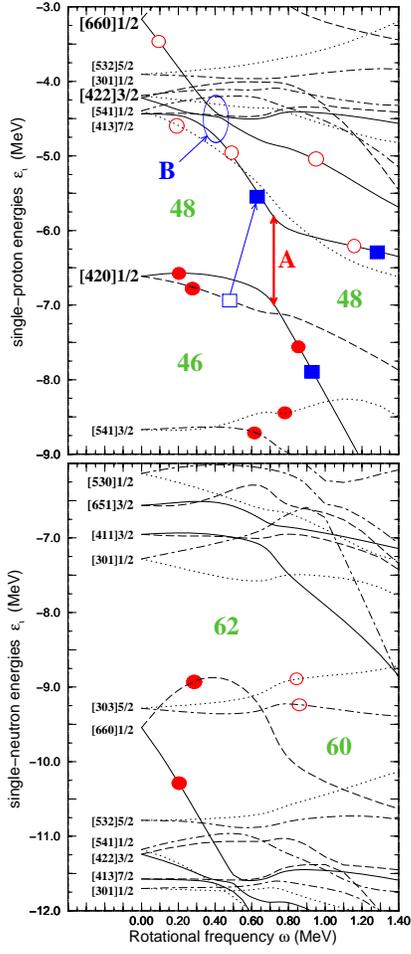}
\caption{\label{routh}
Proton (top) and neutron (bottom) single-particle 
energies (routhians) in the self-consistent rotating potential as a 
function of the rotational frequency $\omega$. They are given along the 
deformation path of the configuration ($\pi 6^1 \nu 6^2$ above
the band crossing) assigned to the band 1. Solid, 
short-dashed, dot-dashed and dotted lines indicate $(\pi=+, r=-i)$, 
$(\pi=+, r=+i)$, $(\pi=-, r=+i)$ and $(\pi=-, r=-i)$ orbitals, 
respectively. At $\omega=0.0$ MeV, the single-particle orbitals are 
labeled by means of the asymptotic quantum numbers $[Nn_z\Lambda]\Omega$ 
(Nilsson quantum numbers) of the dominant component of the wave 
function. Large Nilsson labels are used to indicate the proton 
orbitals which participate in unpaired band crossings seen in the 
bands 1 and 2. These unpaired band crossings are indicated by the 
arrow A (band 1) and ellipse B (band 2). Solid (open) circles indicate 
the orbitals occupied (emptied) in the configuration assigned to the 
band 1. Solid (open) squares show the orbitals occupied (emptied) in 
the configuration of the band 2 as compared with the one of 
the band 1.}
\end{figure}

\begin{figure}[ht]
\includegraphics[angle=0,width=8.0cm]{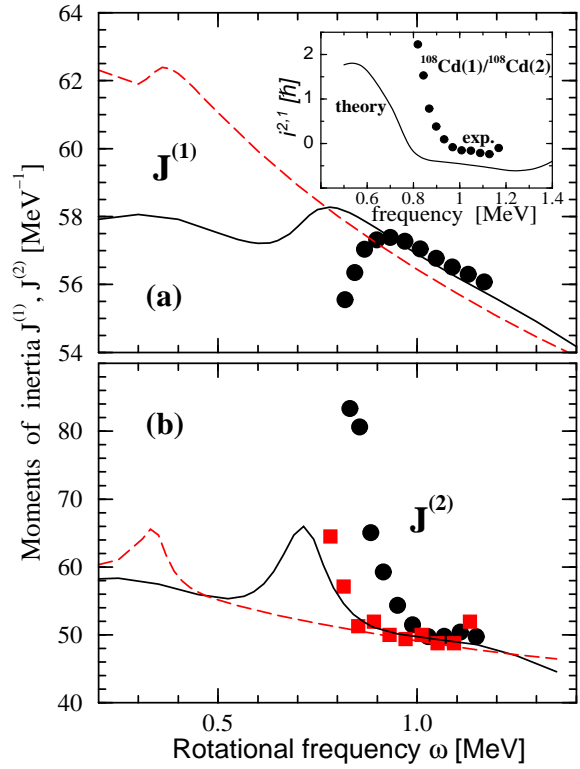}
\caption{\label{j1}
Kinematic (panel (a)) and dynamic (panel (b)) moments of inertia. 
Circles and squares are used for the bands 1 and 2, respectively.
Their theoretical counterparts are shown by solid and dashed lines,
respectively. Kinematic moment of inertia of the band 1 is shown 
under assumption that the spin of its initial state is $I_0=44\hbar$.
The comparison of calculated and experimental kinematic moments
of inertia for band 2 as well as relative alignment 
analysis suggest spin $I_0=43\hbar$ for its lowest state. With 
this spin assignment $J^{(1)}$ moment of inertia of 
band 2 is lower by only $\sim 0.2$ MeV$^{-1}$ than $J^{(1)}$ of 
band 1 at $\omega \geq 0.95$ MeV and, thus, for simplicity it is 
not shown in panel (a). The insert compares calculated and 
experimental (with suggested spin assignments) relative alignments 
of bands 2 and 1, defined as $i^{2,1}(\omega)= 
I_2(\omega)-I_1(\omega)$ \protect\cite{A150}.  }
\end{figure}

 The lowest SD configuration has $(\pi=+, r=+1)$ (even spins) 
quantum numbers and it is assigned to band 1. 
We assign bands 1 and 2 to SD configurations, because the
calculated kinematic $J^{(1)}$ and dynamic $J^{(2)}$ 
moments of inertia agree well with experiment. The configurations with 
normal deformation have too small and the configurations with hyperdeformation 
too large values of moments of inertia. The experimental values of 
transition quadrupole moment $Q_t$ are best reproduced by the 
SD configurations. The proton and neutron 
single-routhian diagrams along the deformation path of this 
configuration are shown in Fig.\ \ref{routh}. Before unpaired band 
crossing it has the $\pi 6^0 \nu 6^2$ structure, while after band 
crossing the $\pi 6^1 \nu 6^2$ structure. The unpaired band 
crossing is due to the crossing of the $\pi [420]1/2(r=-i)$ and  
$\pi [660]1/2(r=-i)$ orbitals (arrow A in Fig.\ \ref{routh}).
The exact band crossing frequencies are not known for bands 1 and 
2 because downsloping branches of $J^{(2)}$ below the band crossing 
have not been observed. The comparison of the experimental 
and calculated bumps in $J^{(2)}$ suggests that the  crossing takes 
place $\approx 200$ keV earlier in the calculations 
than in experiment (see Fig.\ \ref{j1}). 
The frequency of this band crossing depends on the relative 
position of the above mentioned orbitals and thus the discrepancy 
between experiment and theory suggests that the relative energy 
distance between these orbitals is underestimated in the calculations 
by approximately 0.7 MeV. The CRMF calculations well reproduce 
the absolute value of dynamic moment of inertia $J^{(2)}$ above 
the band crossing, but underestimate somewhat the height of the bump 
in dynamic moment of inertia $J^{(2)}$ at the band crossing 
(Fig.\ \ref{j1}b).

 The best agreement between calculated and
'experimental' kinematic moments of inertia $J^{(1)}$ is seen
if the lowest state in band 1 has spin $I_0=44\hbar$
(Fig.\ \ref{j1}a). This suggests that the band in 
$^{108}$Cd has been observed in the spin range from $44\hbar$ 
up to $66\hbar$. However, taking into account that typical accuracy 
of the description of the moments of inertia in the CRMF calculations 
is around 5\% 
\cite{A150,A60,VRAL} and possible minor impact of pairing at high 
rotational frequencies, which would lead to slight decrease of 
calculated $J^{(1)}$ \cite{VRAL}, one cannot completely exclude 
that the lowest SD band in $^{108}$Cd has been  observed from 
$42\hbar$ up to $64\hbar$. The spin range $I=40(2)-60(2)\hbar$
has been suggested in Ref.\ \cite{Cd108-1} using the 
assumption that $J^{(1)}\simeq J^{(2)}$. However, this assumption 
is not supported by our calculations where $J^{(2)}\approx 0.88 
J^{(1)}$ for assigned configuration above the band crossing.

 The average transition quadrupole moment of the assigned configuration 
in the suggested spin range is $Q_t\approx 10.8$ $e$b, which agrees 
with the lower limit of $Q_t=9.5$ $e$b obtained in experiment 
\cite{Cd108-1}. The occupation of the first proton $N=6$ orbital 
at the band crossing has only minor impact on $Q_t$: an increase 
of $Q_t$ by approximately 0.2 $e$b.

     Band 2 has the features similar to the band 1 except
that the band crossing takes place $\approx 200$ keV earlier 
(Fig.\ \ref{j1}b). Especially interesting is the observation that 
the $J^{(2)}$ moments of inertia of these two bands are very 
similar above the band crossing. 
Similar to the $A\sim 150$ region of superdeformation \cite{BRA.88,A150}, 
this suggests that the intruder content of observed bands should 
be the same.  The only possible explanation found 
in our calculations is related to the excitation of proton from 
the $\pi [420]1/2(r=+i)$ orbital into the lowest positive parity 
orbital with signature $r=-i$ above the $Z=48$ shell gap (see Fig.\ 
\ref{routh}). In Fig.\ \ref{routh} above $\omega=0.2$ MeV, this 
orbital has the $\pi [422]3/2$ structure. However, contrary to the 
situation shown in Fig.\ \ref{routh} at equilibrium deformation 
of the configuration of interest the $[422]3/2(r=\pm i)$ orbitals 
are located below the $\pi [413]7/2(r=\pm i)$ orbitals even at
zero rotational frequency. At $\omega \approx 0.35$ MeV, the 
$\pi [422]3/2(r=-i)$ and the $\pi [660]1/2(r=-i)$ orbitals interact
strongly. This interaction creates the bump in 
$J^{(2)}$ (Fig.\ \ref{j1}b). Above the band crossing, the 
calculated $J^{(2)}$ reproduces well the experimental one and
is very close to the $J^{(2)}$ values of the configuration 
assigned to the lowest SD band, the feature observed also
in experiment.

  When interpreting band 2, we also considered particle-hole 
excitations in neutron subsystem, namely from the $\nu[660]1/2(r=+i)$ 
orbital to the $\nu [303]5/2(r=\pm i)$ orbitals (Fig.\ \ref{routh}).
Such excitations lead to smaller deformation, and thus to the delay 
of the unpaired band crossing (originating from interaction of the 
$\pi [420]1/2(r=+i)$ and the $\pi [660]1/2(r=+i)$ orbitals).
For example, in the configuration based on the occupied 
$\nu [303]5/2(r=-i)$ orbital this crossing is delayed by 
$\approx 0.35$ MeV, which  is in contradiction with 
experiment.

  Therefore, similar to the configuration assigned to the 
band 1, the structure of the configuration assigned to the band 2
changes from $\pi 6^0 \nu 6^2$ to $\pi 6^1 \nu 6^2$ at the 
band crossing. However, it has total signature $r=-1$ (odd spins)
and negative parity. 
The band crossing A, active in the configuration of the band 1, 
is not active in the configuration of band 2 since both interacting 
orbitals ($\pi [420]1/2(r=-i)$ and $\pi [660]1/2(r=-i)$) are occupied (see 
Fig.\ \ref{routh}). Since the $\pi [413]7/2(r=-i)$ and $\pi [660]1/2(r=-i)$ 
orbitals interact very weakly (see inside of ellipse B in Fig.\ 
\ref{routh}), their interaction cannot explain the observed features 
of band 2 in the band crossing region.
 The comparison of relative alignments of bands 2 and 1 
(Fig.\ \ref{j1}) suggests that with $I_0=44\hbar$ assigned for initial 
state of band 1, the band 2 has been observed in the spin range from 
$I_0=43\hbar$ up to $I=65\hbar$. In the region beyond band crossing 
($\omega=0.9-1.2$ MeV),  the calculated relative alignments differ from 
the experimental ones only by $\approx 0.3\hbar$: this is within the 
typical uncertainty of the description of relative alignments in the 
CRMF theory \cite{A150-sp,A60,A80}.

   The calculations show that the quadrupole deformations of the 
configurations assigned to the bands 1 and 2 are almost the
same, but they differ somewhat in the $\gamma$-deformation.
While the configuration of band 1 has $\gamma \approx 6^{\circ}$, 
the one of the band 2 has $\gamma \approx -4^{\circ}$. As a 
consequence, the configuration of the band 2 is more collective 
($Q_t\approx 12.2$ $e$b) as compared with the one of band 1. Although
the experimental analysis for band 2 yielded a lower
value for the transition quadrupole moment $Q_t=8.5$ $e$b,
it suffered from the large errors on the $F(\tau)$ values 
which did not allow an accurate measurement of the $Q_t$ 
values \cite{Cd108-2}. Thus based on the similarity of
rotational properties of bands 1 and 2 above the band crossing 
and the results of the CRMF calculations, it is reasonable to 
believe that the deformation for band 2 is comparable to the 
one for band 1 (see also Ref.\ \cite{Cd108-2}).

  The earlier band crossing in the configuration of band 2 as 
compared with experiment is due to small energy distance between the 
$\pi [422]3/2$ and $\pi [660]1/2$ orbitals in the calculations. 
In order to have this crossing at experimentally observed 
frequency, one should increase this distance by $\approx 0.7$ 
MeV. This value is similar to the underestimate of the energy
distance between the $\pi [420]1/2$ and $\pi[660]1/2$ orbitals 
deduced earlier.
This suggests that the energy distance between the $\pi [422]3/2$
and $\pi [420]1/2$ orbitals, and, consequently the $Z=48$ SD shell 
gap is reasonably well reproduced in the calculations.
Thus, one concludes that the $Z=48$ SD shell gap is smaller 
than the $Z=30$ SD shell gap in the $A\sim 60$ region of 
superdeformation (see Fig.\ 1 in Ref.\ \cite{A60}) and 
$Z=66$ SD shell gap in the rare-earth region of 
superdeformation (see Fig.\ 3 in Ref.\ \cite{A150}). 
This maybe a possible reason why the search for SD bands
in neighbouring nuclei has been unsuccesful so far 
\cite{F.05}. 

  In addition, it indicates that in reality the $\pi[660]1/2$ 
orbital is located  $\approx 0.7$ MeV higher in energy with 
respect of the $\pi [420]1/2$ and $\pi [422]3/2$ orbitals than 
suggested by the present calculations. This is not far away 
from the typical accuracy with which the energies of the 
single-particle states are reproduced at normal- \cite{A250} 
and superdeformation \cite{A150-sp} in the RMF calculations 
with the NL1 set.

  Unpaired band crossings at superdeformation were reported
earlier, for example, in $^{146,147}$Gd \cite{H.93,A150} 
and $^{153}$Ho \cite{Ho153}. The band crossing in the SD band 
1 of $^{153}$Ho is similar to the ones seen in $^{108}$Cd since
it involves the interaction of the routhians with 
$\Delta N=2$.

\begin{figure}[h]
\includegraphics[angle=0,width=7.0cm]{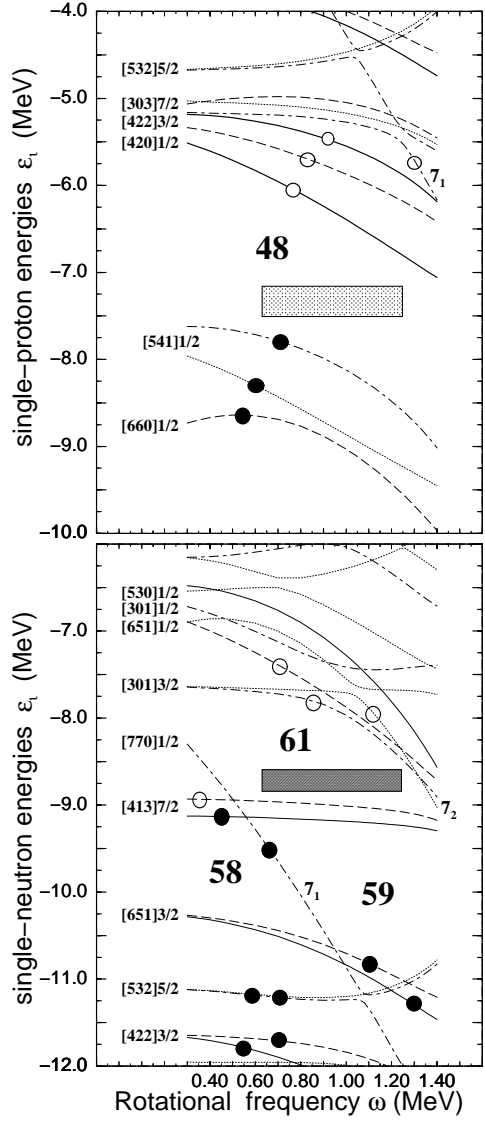}
\caption{\label{routh-HD} The same as Fig.\ \protect\ref{routh},
but along the deformation path of hyperdeformed $\pi 6^2 
\nu 6^4 7^1$ configuration. Dashed box indicates the frequency 
range corresponding to the spin range $I=44-84\hbar$ in this
configuration. Solid (open) circles indicate  the occupied 
(emptied) orbitals. The labels $7_{1,2}$ are used to show
two lowest $N=7$ orbitals.}
\end{figure}

\begin{figure}[ht]
\includegraphics[angle=0,width=8.0cm]{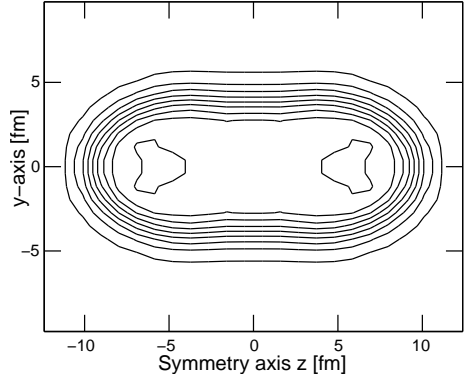}
\caption{\label{density} 
 The self-consistent neutron density $\rho_n(y,z)$ as a function 
of $y-$ and $z-$ coordinates for lowest HD configuration 
($\pi 6^2 \nu 6^4 7^1$) at rotational frequency $\omega=1.0$ MeV. 
The densities are shown in steps of 0.01 fm$^{-3}$ starting from 
$\rho_n(y,z)=0.005$ fm$^{-3}$.
}
\end{figure}

   The hyperdeformed configurations become yrast above $I=68\hbar$
(Fig.\ \ref{eld}). These are the signature partner $\pi 6^2 \nu 6^4 7^1$ 
configurations of negative parity which are degenerated in energy. 
This energy degeneracy is due to the excitation of one neutron from the 
$\nu [413]7/2(r=\pm i)$ orbitals into the lowest $N=7$ orbital (see 
Fig.\ \ref{routh-HD}). The HD configurations are favoured in energy 
at these spins 
due to the $Z=48$ HD shell gap and low neutron level density at 
$N=59-61$. The transition quadrupole moment $Q_t$ increases
from $Q_t=17.2$ $e$b at $I=44\hbar$ up to $Q_t\approx 17.5 \hbar$
at $I=84\hbar$. This fact is related to the stretching of the nucleus 
due to centrifugal force. The shape of nucleus corresponds to the
2.3:1 semiaxis ratio (Fig.\ \ref{density}). The kinematic moment of 
inertia drops down from $J^{(1)}\approx 70.5$ MeV$^{-1}$ at $I=44\hbar$ to 
$J^{(1)}\approx 67.5$ MeV$^{-1}$ at $I=84\hbar$. In the frequency 
range where the experimental bands 1 and 2 have been observed, the 
dynamic moments of inertia of calculated configurations are equal 
to $J^{(2)}\approx 64$ MeV$^{-1}$; the quantity which is much 
larger than in experiment. At higher frequencies there is smooth 
increase of $J^{(2)}$ up to $\approx 70$ MeV$^{-1}$ at 
$\omega \sim 1.4$ MeV, possibly caused by unpaired band 
interaction. No experimental counterparts of these configurations 
have been observed so far.

  It is interesting to mention the similarity of the single-particle 
spectra at hyperdeformation obtained in the CRMF calculations (Fig.\ 
\ref{routh-HD}) and in the Wood-Saxon-Strutinsky calculations of 
Ref.\ \cite{C.01} (Fig.\ 4 and 5 of Ref.\ \cite{C.01}). However,
the $N=61$ ($N=58$ and $N=59$) shell gaps are somewhat smaller
(larger) in the CRMF calculations. In the later calculations, the 
quadrupole, octupole, hexadecapole and necking degrees of freedom 
were taken into account. There are, however, no indications that
octupole deformation should play a role at HD in $^{108}$Cd 
\cite{C.01}. On the other side, the CRMF calculations take into 
account all even-multipole deformations and triaxiality
in a fully self-consistent way. The basis of the CRMF model is 
sufficiently large to see if there is a pronounced tendency for 
the development of necking. Fig.\ \ref{density}, showing the 
densities of lowest HD configuration, does not indicate such a  
tendency.  

  The CNS calculations, which will not be discussed here in detail, 
also indicate that the band 1 in $^{108}$Cd is associated with the 
$\pi 6^1 \nu 6^2$ configuration having $Q_t \sim 10.3 $ $e$b and 
located  in the $\varepsilon_2 \sim 0.67, \gamma \sim 15^{\circ}$
minimum. Similar to earlier calculations (Ref.\ \cite{A60}), the 
triaxiality is more pronounced in the CNS calculations for 
$^{108}$Cd as compared with the CRMF ones. In the CNS calculations, 
the HD configurations are yrast above $I \approx 63\hbar$, and the 
$\pi 6^2 \nu 6^4 7^1$ configurations with $Q_t \approx 18$ $e$b are 
the lowest ones. Thus, one can conclude that general properties of 
SD and HD bands are similar in both approaches.

   The present CRMF calculations suggest that in the spin range 
where the feeding of SD bands takes place (i) the SD bands are 
non-yrast and (ii) the HD bands compete in energy with the SD 
bands. However, only SD bands have been reported at high spin 
in Refs.\ \cite{Cd108-1,Cd108-2}. Because of limited 
experimental information it is difficult to find unique explanation 
for this discrepancy. On one side, one would expect that the 
calculated relative energies of different minima deviate somewhat 
from reality. Alternative explanation is related to the differences
in feeding mechanism of SD and HD bands \cite{Dudek}. Although 
the HD and SD states are comparable in energy at $I\sim 66 
\hbar$ at zero temperature (Fig.\ \ref{eld}), these high spin 
nuclear states are populated at somewhat higher spin at 
temperatures $T \sim 1.0$ MeV. At these temperatures, the 
deformed shell structure melts, and thus SD minimum becomes 
lower in energy than HD one.  As a consequence, the feeding 
of the SD minimum maybe the dominant population channel 
\cite{Dudek}.

    In summary, the detailed analysis of the structure of 
observed superdeformed bands 1 and 2 in $^{108}$Cd within
the cranked relativistic mean field theory led to new 
configuration assignment: both bands have been assigned the
$\pi 6^1 \nu 6^2$ configurations. The band crossings seen in 
these bands have been related to the unpaired crossings of the lowest 
proton $N=6$ orbital and the $N=4$ orbitals. The size of the $Z=48$
SD shell gap, which is smaller than that of doubly magic $^{60}$Zn 
and $^{152}$Dy SD nuclei, and the absence of large neutron shell 
gap at $N=60$ indicate that the $^{108}$Cd nucleus cannot be considered 
as a doubly magic SD nucleus. The properties of hyperdeformed 
configurations, which are expected to be yrast above $I\approx 64\hbar$, 
have also been discussed.

\section{Acknowledgements}

  The work was supported by the DoE grant DE-F05-96ER-40983. The help
of Y.\ Gu in performing CNS calculations and the discussions with 
I.\ Ragnarsson are highly appreciated.

\vspace{0.5cm}

\end{document}